\documentclass[11pt,a4paper]{article}
\pdfoutput=1
\usepackage{jheppub}

\usepackage{etoolbox}
    \makeatletter
    \patchcmd{\maketitle}{\@fpheader}{}{}{}
    \makeatother

\usepackage{subfigure,amsmath,amssymb ,amsfonts, latexsym}

\usepackage[notcite,color,final]{showkeys}
\definecolor{refkey}{gray}{.25}
\definecolor{labelkey}{gray}{.25}

\newcommand{\be}{\begin{eqnarray}}
\newcommand{\ee}{\end{eqnarray}}

\def\ben{\begin{equation}}
\def\een{\end{equation}}
\def\bena{\begin{eqnarray}}
\def\eena{\end{eqnarray}}

\newcommand{\beq}{\begin{equation}}
\newcommand{\eeq}{\end{equation}}
\newcommand{\bea}{\begin{eqnarray}}
\newcommand{\eea}{\end{eqnarray}}

\newcommand{\bean}{\begin{eqnarray*}}
\newcommand{\eean}{\end{eqnarray*}}
\newcommand{\bs}{\begin{subequations}}
\newcommand{\es}{\end{subequations}}

   \vspace*{0cm}

\title{Emerging Potentials in Higher-Derivative Gauged Chiral Models Coupled to ${\cal{N}}=1$ Supergravity}

\author{Fotis Farakos and}
 \author{Alex Kehagias}

\affiliation{Physics Division, National Technical University of Athens,
15780 Zografou Campus,  Athens, Greece}

\emailAdd{fotisf@mail.ntua.gr}
\emailAdd{kehagias@central.ntua.gr}


\abstract{We present a new method to introduce scalar potentials to gauge-invariant chiral models 
coupled to supergravity. The theories under consideration 
 contain consistent higher-derivative terms which do not give rise to instabilities and 
ghost states. The chiral auxiliaries are  not propagating and  can be integrated out. Their elimination
gives rise to emerging potentials even when there is not a superpotential to start with. We present the case of a single 
chiral multiplet with and without a superpotential and, in the gauged theory, up to two chiral multiplets coupled to supergravity 
with no superpotential. A general feature of the emergent potential is that it is negative defined leading to 
anti-de Sitter vacua. In the gauge models, competing D-terms may lift the potential leading to stable and 
metastable de Sitter and Minkowski vacua as well with spontaneously broken supersymmetry.
}

\begin{document}

\maketitle

\section{Introduction and Conclusions}

Higher derivative theories are usually ill-defined in the sense that they  suffer from  Ostrogradski instabilities 
\cite{ostro}. These are
instabilities caused by terms linear in momentum  in the Hamiltonian as a result of the higher time derivatives. 
Quantum mechanically
these instabilities are usually  shown up as ghost states. It is important therefore to determine what kind   of
higher-derivative interactions may give rise to consistent theories.  
In the case of scalars coupled to gravity for example, the consistent higher 
derivative theories we know, (which are also quadratic in the scalars)  contain the terms   
\be
 {\cal{L}}_{I\phantom{I}}&=&\phi^2 R_{GB}^2\, ,\label{GB}\\
{\cal{L}}_{II}&=& G^{\mu\nu}\partial_\mu \phi\partial_\nu \phi\, , \label{Gf}
\ee
where
\be
G_{\mu\nu}=R_{\mu\nu}-\frac{1}{2} g_{\mu\nu}R\, , ~~~ R_{GB}^2=R_{\mu\nu\gamma\delta}R^{\mu\nu\gamma\delta}
-4 R_{\mu\nu} R^{\mu\nu}+R^2
\ee   
are the Einstein  and the Gauss-Bonnet tensors, respectively.
 When more fields are involved, the general case has been discussed in 
\cite{hord}, whereas  such theories are nowadays known as galileon \cite{gal2,gal3}, or k-essence, for example \cite{dm, m2,m1}.

That ${\cal{L}}_{I\phantom{I}}$ leads to second order evolution
equation follows easily from the fact that the Gauss-Bonnet combination is a total derivative in four-dimensions and it is 
linear in second order derivatives. Instead, ${\cal{L}}_{II}$ leads to second order equations as, in Hamiltonian ADM 
formalism \cite{adm}, $G_{tt}$ and $G_{it}$ contain only first time
derivatives, as $G_{tt}$ and $G_{ti}$ are the Hamiltonian and momentum constraints.

In a supergravity setup now, the supersymmetrization of ${\cal L}_{I}$  has been worked out in \cite{fer1, fer2} and of
${\cal L}_{II}$ in \cite{ger-far}. 
Here, we construct higher-derivative theories of k-essence type, i.e. theories that 
the scalar kinetic term $K(X)$ is in general a function of $X=(\partial\phi)^2$ \cite{dm, m2,m1}. Such theories    
apart from their obvious interest as the 
most generic supersymmetric theories avoiding Ostrogradski (higher derivatives) 
instabilities, may also have phenomenological interest. 

It should be stressed that in general there are two different types of higher derivative interactions: those which lead to 
equations of motions with higher than two time derivatives and those which contains at most two time  derivatives. The first class
leads at the classical level to  the  Ostrogradski instability and at the quantum level to ghost states. Here we are interested only 
in the second class of models where there are no higher that two time derivatives in the field equations. We 
also note that the higher derivative terms we are employing here are not new (they have appeared at different occasions for different reasons).
Such terms have been considered before in global supersymmetry in
\cite{gates,ovrut1,Khoury:2010gb}, in which cases the building block for the theory we consider was introduced. 
In particular, such term has been employed in a generic contruction of ${\cal{N}}=1$ globally supersymmetric
Gallileon \cite{ovrut1} and k-essence type lagrangians \cite{Khoury:2010gb}. Motivated by the latter constructions, 
and our earlier work in the new minimal supergravity framework \cite{ger-far},  one is tempted to 
consider such higher derivative interactions in a locally supersymmetric setup.
Although we do not have a recipe 
of which higher derivative interactions do not lead to higher time derivatives, the particular form of the 
higher derivative terms we will employ satisfy this criterion and therefore are free from Ostrogradski instabilities and ghost 
states. 
Just before the appearence
of this work, another article
with the same higher derivative interactions appeared in \cite{LO} discussing however, the ungauged case with 
a single chiral multiplet.

In our work, we are focusing on the emergence of the scalar potential for more than 
one (two) chiral multiplets in general {\it gauged} supergravity models. A generic feature of these models
is the modification of the scalar potential.  
The fact that higher derivative models indeed modify the effective scalar potential is known. 
For example, \cite{Cecotti:1986jy}
dicuss the structure of the scalar potential in general higher derivative theories. 
In this work, we construct models which explicitly realise this idea. 

There are various reasons for which one should consider the most general type of potential in supersymmetric theories.
At a more theoretical level, 
one fundamental question is if the structure of the standard scalar potential at the two derivative level is maintained when higher
derivatives are included. We find, consistently with previous suggestions \cite{Cecotti:1986jy}, 
that the scalar potential has completely different 
structure in the presence of higher derivatives. This has profound effects to the structure of the vacuum and the 
 supersymmetry breaking.
Also, the gauge models
we are discussing (and which have not been employed previously) could lead to different Higgs sectors and potential as the two chiral multiplets 
we are considering could be the Higgs mutiplets. In addition, as the theory we will describe does not contain extra states besides the usual
graviton and the scalars (in the bosonic sector), it can be considered as inflaton (or curvatons) in   cosmology. In this context, 
density perturbations of the scalar fields  will then be a issue to be explored.
Moreover, this new scalar potential seems to have very interesting properties from the supersymmetry breaking point of view, since it breaks it spontaneously. Thus it can provide new possibilities in a hidden sector supersymmetry breaking setup.

We will consider chiral multiplets coupled to supergravity. We will write down supersymmetric actions in ${\cal{N}}=1$ superspace
and  explicitly work out the case of a single chiral multiplet without superpotential. We will see that in this case, the 
action contains quartic powers of the auxiliaries which, nevertheless, can  still be eliminated by their (algebraic) equations of motions.
As a result of the elimination of the chiral auxiliaries, a potential for the scalar emerges even though there 
is no superpotential to start with.
Interesting enough, supersymmetry is always spontaneously broken and there is no supersymmetric vacuum. The same can be done when
the chiral multiplet possess a superpotential as well. In this case, although more involved, the auxiliaries can still be eliminated 
giving rise to a quite complicated scalar potential, completely changing the standard ${\cal{N}}=1$ one. I addition, we exploit the case of  more than one chiral multiplets coupled to 
supergravity. Although in this case, the problem is technically more difficult as one has to solve algebraic equations of 
increasing higher order as we increase the number of chirals , 
we do explicitly a two-chiral multiplets model with no superpotential.

We consider gauge invariant chiral models as well, where some of the isometries of the scalar manifold are gauged. In particular we discuss
a single chiral multiplet charged under some Abelian gauge group and no superpotential. In this case, on top of the 
D-term contribution to the scalar potential, there exists a new contribution coming from the emergent potential after integrating 
out the scalar auxiliaries. We also introduce FI-terms and we find the saddle points of the   potential,
which may possess de Sitter, Minkowski or anti-de Sitter symmetry. Again supersymmetry is always spontaneously 
broken and there is no supersymmetric vacuum.   
Finally, we also discuss gauge models with two chiral charged multiplets and no-superpotential. This case can also be solved 
exactly with similar results.     

The lesson from these considerations is that higher derivative terms in supergravity contribute also to the potential terms. Thus, 
theories with no potential at the leading two-derivative level, may develop nontrivial potential when higher derivatives are 
taken into account. At this point there are however, two dangerous aspects. The first one concerns the appearance of the Ostrogradski
instability. In the type of theories we are discussing, this instability is not present as the theory although with 
higher derivatives, gives rise to equations of motion with maximum second order time derivatives. 
This guarantees causal propagation and no-unphysical 
ghost states. The second issue concerns the auxiliary fields. Here, we were able still to eliminate the auxiliaries of the 
chiral multiplet since they appeared algebraically in the supersymmetric Lagrangian. In fact, only the modulus of the auxiliary 
could be eliminated while its phase is propagating. This is enough however, as in the bosonic action, only its modulus appear
and therefore can be integrated out. It may happen, and this is the rule actually, that in supersymmetric theories 
with higher-derivatives, the auxiliaries turns out to be dynamical and propagating. In that case, they can not 
be eliminated and they should be kept in the total Lagrangian. Under such circumstances there is no emergent  potential.

Summarizing, we stress that the standard form of the supersymmetric potential for chiral fields coupled to ${\cal{N}}=1$
supergravity
\be
\label{poten} 
V= e^{K} \left[ g^{i {\bar{j}}} (D_iP) (D_{\bar{j}}\bar{P}) -3 P \bar{P} \right]
\ee
is only valid at the leading two-derivative level. Introduction of appropriate higher derivative terms that  
do not give rise to pathologies and inconsistencies, modifies considerably the structure of (\ref{poten}) 
and give rise to a kind of emergent potential even in cases where there is no one to start with (i.e. no superpotential).
In this case, supersymmetry is always spontaneous broken with de Sitter, Minkowski or anti-de Sitter maximally 
symmetric vacua. This is also true in the case of gauged models with or without FI-terms.

\vskip.1in

In the next  section 2. we present the higher-derivative chiral model coupled to 
${\cal{N}}=1$ supergravity. In section 3. we  discuss the emergent potential of a single chiral multiplet 
with and without superpotential and in section 4. we present the emergent potentials for gauge models.

\section{Chiral Models with Higher Derivatives in Supergravity}

Let us consider the most general (two-derivative) superspace Lagrangian of chiral superfields
 coupled to supergravity in superspace formalism
\footnote{Our framework  and  conventions are those of Wess and Bagger \cite{Wess:1992cp}.}
\beq
\label{W&B}
{\cal{L}}_0=\frac{1}{\kappa^2} \int d^2 \Theta \, \,2 {\cal{E}} \left[ \frac{3}{8} \Big{(} \bar{{\cal{D}}}\bar{{\cal{D}}} -
 8 {\cal{R}} \Big{)}\,  e^{ - \frac{\kappa^2}{3} K(\Phi^i,\bar{\Phi}^{\bar{j}})} +\kappa^2 P(\Phi) \right] + h.c.
\eeq
The hermitian function $K(\Phi^i,\bar{\Phi}^{\bar{j}})$ is the K\"ahler potential \cite{Zumino:1979et}, 
$P(\Phi^i)$ is the superpotential (a holomorphic function of the chiral superfields $\Phi^i$) and $\kappa$ is 
proportional to the Planck length, which from now on will be set equal to 1. 
The abbreviations 
\be
\bar{{\cal{D}}}_{\dot{\alpha}}\bar{{\cal{D}}}^{\dot{\alpha}}=\bar{{\cal{D}}}\bar{{\cal{D}}}\, , ~~~~~
{\cal{D}}^{\alpha}{\cal{D}}_{\alpha}={\cal{D}}{\cal{D}}
\ee
 will be used for 
the sum of the fermionic superspace covariant derivatives and $\int d^{2} \Theta$ is the super-integration over the so
 called new $\Theta$ variables. From the supergravity multiplet sector, $2{\cal{E}}$ is the usual chiral density employed to
 create supersymmetric Lagrangians, which  in the new $\Theta$ variables has the expansion
\beq
\label{2Epsilon}
2{\cal{E}}=e \left\{ 1+ i\Theta \sigma^{a} \bar{\psi}_a -\Theta \Theta\Big{(} M^* +\bar{\psi}_a\bar{\sigma}^{ab}\bar{\psi}_b 
\Big{)} \right\}
\eeq
in terms of  the vielbein ($e_m^a$), the gravitino ($\psi_m$) and 
 the complex scalar auxiliary field $M$. We also mention that the off-shell minimal supergravity multiplet also 
contains another auxiliary field, the real vector $b_a$. In addition, ${\cal{R}}$, the superspace curvature, is
 a chiral superfield which contains the Ricci scalar in its highest component. 
In the matter sector, $\Phi^i$ and $\bar{\Phi}^{\bar{j}}$ denote a set on chiral and anti-chiral superfields ($\bar{{\cal{D}}}_{\dot{\alpha}}\Phi^i=0 $, ${\cal{D}}_{\alpha}\bar{\Phi}^{\bar{j}}=0 $) whose components are defined via projection
\bea
\label{chiral-def}
\nonumber
A^i&=&\Phi^i|_{\theta=\bar{\theta}=0}, \\
\chi^i_\alpha&=&\frac{1}{\sqrt{2}} {\cal{D}}_{\alpha} \Phi^i |_{\theta=\bar{\theta}=0}, \\
\nonumber
F^i&=&- \frac{1}{4}  {\cal{D}}{\cal{D}} \Phi^i|_{\theta=\bar{\theta}=0}.
\eea 
After calculating the component form of (\ref{W&B}), integrating out the auxiliary fields and performing a Weyl rescaling of the gravitational field (accompanied by a redefinition of the fermionic fields), the pure bosonic Lagrangian reads
\beq
\label{W&B-OnShell-rescaled}
e^{-1} {\cal{L}}_0=-\frac{1}{2} R - g_{i {\bar{j}}} \partial_a A^i \partial^a \bar{A}^{\bar{j}} - \ e^{K} \left[ g^{i {\bar{j}}} (D_iP) (D_{\bar{j}}\bar{P}) -3 P \bar{P}   \right].
\eeq
Further details maybe found for example in \cite{Wess:1992cp}. Here 
\be
g_{i {\bar{j}}} = \frac{\partial^2 K(A,\bar{A})}{\partial A^i \partial\bar{A}^{\bar{r}}}
\ee
 is the positive definite K\"ahler metric, on the manifold parametrized by $A^i$ and $\bar{A}^{\bar{j}}$. 
Moreover, the K\"ahler space covariant derivatives are defined as follows
\beq
\label{covariant-Khaler}
D_iP=P_i +K_iP
\eeq
where 
\be
P_i=\frac{\partial P }{ \partial A^i} \, , ~~~~K_i=\frac{\partial K }{ \partial A^i}\, .
\ee 
The Lagrangian (\ref{W&B-OnShell-rescaled}) is K\"ahler invariant as long as the superpotential  scales as 
\be
P(A^i)\rightarrow e^{-S(A^i)} P(A^i)
\ee
 under a K\"ahler transformation 
\be
K(A^i,\bar{A}^{\bar{j}})\rightarrow K(A^i,\bar{A}^{\bar{j}})+S(A^i)+\bar{S}(\bar{A}^{\bar{j}}),
\ee 
where $S(A^i)$ and $\bar{S}(\bar{A}^{\bar{j}})$ are holomorphic functions of the complex coordinates. 
From the form of the supersymmetry transformations of the fermions one can 
see that
\be
\delta_{susy} \chi \sim F
\ee
Thus, supersymmetry is broken whenever 
\be
 <D_{i}P> \ne 0
\ee
 since on-shell 
\be
F^* \sim -\frac{\partial P}{\partial A} - P \frac{\partial K}{\partial A} = D_{A}P\, .
\ee
This fact leads to the conclusion that it is possible to have supersymmetric anti-de Sitter (AdS) or Minkowski vacua but 
not de Sitter (dS) ones since $-3 P \bar{P} \le 0$ always. 
It should be stressed that this is a property of the superpotential and not a general property of the 
supergravity theory after integrating out the auxiliary sector. Indeed, there are cases where anti-de Sitter vaccua maybe
uplifted   to de Sitter ones \cite{Kachru:2003aw,Villadoro:2005yq,Catino:2011mu}.

\subsection{Higher Derivative Chiral Models}

Higher derivative couplings have extensively been studied. Nevertheless, not all possible such higher derivative 
terms have an exact supergravity counterpart, and some might not have one at all. 
The theory we are interested in, has a superspace Lagrangian of the form 
\be
{\cal{L}}={\cal{L}}_0+{\cal{L}}_{HD} \label{lang}
\ee
where ${\cal{L}}_0$ is the standard superspace supergravity Lagrangian given in eq.(\ref{W&B}) and 
\beq
\label{KSB-Kahler-inv}
{\cal{L}}_{HD}=\int d^2 \Theta \, \, 2 {\cal{E}} \left\{ \frac{1}{8} 
\Big{( }\bar{{\cal{D}}}\bar{{\cal{D}}} - 8 {\cal{R}} \Big{)}   
\Lambda^{\bar{r}i\bar{n}j} 
\Big{[} 
\bar{{\cal{D}}}_{\dot{\alpha}} K_{i} 
{\cal{D}}_{\alpha}  K_{\bar{r}} 
\bar{{\cal{D}}}^{\dot{\alpha}}K_{j} 
{\cal{D}}^{\alpha}  K_{\bar{n}} 
\Big{]}  \right\} + h.c.
\eeq
Note that (\ref{KSB-Kahler-inv}) appeared first in \cite{LO} just before present work released wheras 
the  flat-space limit of (\ref{KSB-Kahler-inv}) with global supersymmetry appeared in \cite{gates,Khoury:2010gb}.
It is important that ${\cal{L}}$ is manifestly both K\"ahler and (independently) super-Weyl invariant, as will be seen later on. 
These two symmetry properties, although obviously they do not specify the form of the action, they are essential
in the consistency of the model as well as for the supergravity theory that it describes. 
As we will see, (\ref{KSB-Kahler-inv}) does not involve derivatives of the auxiliary fields, which are not propagating and can be integrated out. 
Equivalently, (\ref{KSB-Kahler-inv}) can be expressed in terms of the chiral superfields $\Phi^i$ as
\beq
\label{KSB1}
{\cal{L}}_{HD}=\int d^2 \Theta \, \, 2 {\cal{E}} \left\{  \frac{1}{8} 
\Big{(} \bar{{\cal{D}}}\bar{{\cal{D}}} - 8 {\cal{R}} \Big{)}   \Lambda_{i\bar{r}j\bar{n}} \Big{[}
\bar{{\cal{D}}}_{\dot{\alpha}} {\bar{\Phi}}^{\bar{r}} {\cal{D}}_{\alpha} \Phi^i  \bar{{\cal{D}}}^{\dot{\alpha}}{\bar{\Phi}}^{\bar{n}} 
 {\cal{D}}^{\alpha} \Phi^j \Big{]}  \right\} + h.c.
\eeq
where 
\be
K_{i\bar{r}}= \frac{\partial^2 K(\Phi,\bar{\Phi})}{\partial\Phi^i \partial\bar{\Phi}^{\bar{r}} }
\ee
 is the K\"ahler metric on the complex space spanned by the chiral and anti-chiral superfields and 
$\Lambda_{i\bar{r}j\bar{n}}$ represents a K\"ahler tensor. For example, one may choose
\beq
\label{Lambda-ex}
{\Lambda}_{i\bar{r}j\bar{n}}= {\cal{G}}(\Phi,\bar{\Phi}) K_{i\bar{r}}K_{j\bar{n}}+ {\cal{H}}(\Phi,\bar{\Phi}) {\cal{R}}_{i\bar{r}j\bar{n}}
\eeq
with ${\cal{G}}(\Phi,\bar{\Phi})$ and ${\cal{H}}(\Phi,\bar{\Phi})$ being some K\"ahler invariant hermitian functions and ${\cal{R}}_{i\bar{r}j\bar{n}}$ the K\"ahler space Riemann tensor defined as
\beq
\label{Kahler-R}
{\cal{R}}_{i\bar{j}k\bar{l}}=\frac{\partial}{\partial {\Phi}^i}\frac{\partial}{\partial {\bar{\Phi}}^{\bar{j}}} K_{k\bar{l}} - K^{m\bar{n}} 
\left( \frac{\partial}{\partial {\bar{\Phi}}^{\bar{j}}} K_{m\bar{l}} \right) 
\left( \frac{\partial}{\partial {\Phi}^i} K_{k\bar{n}} \right).
\eeq
The form (\ref{Lambda-ex}) implies some symmetries for the K\"ahler indices which, 
without loss of further generality, we will assume to be possessed by all the 
$\Lambda_{i\bar{r}j\bar{n}}$ to be considered in this work. 
Our next task is to extract the component 
field expression for the  Lagrangian (\ref{KSB1}), which 
after  superspace integration turns out to be
\bea
\label{KSB-comp}
\nonumber
e^{-1}{\cal{L}}_{HD}=-16\ {\cal{U}}_{i\bar{r}j\bar{n}}\!\!
&\!\!&\!\!
\Big{(}F^{i} F^{j}\bar{F}^{\bar{r}}\bar{F}^{\bar{n}}
+\partial_a A^{i} \partial^a A^{j} \partial_b \bar{A}^{\bar{r}} \partial^b \bar{A}^{\bar{n}}
\\
&&
-F^{i} \bar{F}^{\bar{r}} \partial_a A^{j}\partial^a \bar{A}^{\bar{n}}
-F^{i} \bar{F}^{\bar{n}} \partial_a A^{j}\partial^a \bar{A}^{\bar{r}}\Big{)}.
\eea
for the pure bosonic sector. In (\ref{KSB-comp}) we have used the notation
\be
\label{LL}
{\cal{U}}_{i\bar{r}j\bar{n}}(A,\bar{A})=\Lambda_{i\bar{r}j\bar{n}}(\Phi,\bar{\Phi})\Big{|}_{\theta=\bar{\theta}=0}
\ee
Again it is easy to see that (\ref{KSB-comp}) is manifestly K\"ahler invariant.

\subsection{Super-Weyl Invariance}

At this point its is crucial to make a comment on a subtlety concerning the hermitian vector superfield 
\beq
\label{thevector}
V=\Lambda_{i\bar{r}j\bar{n}} \bar{{\cal{D}}}_{\dot{\alpha}} {\bar{\Phi}}^{\bar{r}} {\cal{D}}_{\alpha} \Phi^i 
 \bar{{\cal{D}}}^{\dot{\alpha}}{\bar{\Phi}}^{\bar{n}} {\cal{D}}^{\alpha} \Phi^j ,
\eeq
namely, its scaling properties under super-Weyl transformations.
We emphasize that $V$ is defined through its components. For example, its lowest component will be
\be
\label{lowest-comp}
V\Big{|}_{\theta=\bar{\theta}=0}&=& 4\ {\cal{U}}_{i\bar{r}j\bar{n}}\ \bar{\chi}_{\dot{\alpha}}^{\bar{r}} \chi_{\alpha}^i 
 \bar{\chi}^{\dot{\alpha}\bar{n}} \chi^{{\alpha}j}\label{v}.
\ee
Moreover, all components of $V$ should be understood as those of a hermitian vector superfield 
defined via projection and will eventually be related to (\ref{chiral-def}). 
This definition will gives Weyl weight $-2$ to the vector superfield $V$, as is required so that (\ref{KSB1}) is 
indeed K\"ahler and super-Weyl invariant. These symmetries are crucial for consistency of the supergravity Lagrangian 
 on curved superspace. Under a super-Weyl transformation,
 the superspace covariant derivatives change as \cite{Dragon:1986ew}
\bea
\label{weyl-superderiv} \nonumber
\delta_{\Sigma}{\cal{D}}_{\alpha}=(\Sigma -2 \bar{\Sigma}){\cal{D}}_{\alpha} - ({\cal{D}}^{\gamma} \Sigma )l_{\alpha\gamma} \\
\delta_{\Sigma}\bar{\cal{D}}_{\dot{\alpha}}=(\bar{\Sigma} -2 \Sigma)\bar{\cal{D}}_{\dot{\alpha}} - (\bar{\cal{D}}^{\dot{\gamma}} \bar\Sigma )l_{\dot{\alpha}\dot{\gamma}}
\eea
where the $l_{\alpha\gamma}$ and $l_{\dot{\alpha}\dot{\gamma}}$ stand for the (anti)self-dual parts of 
infinitesimal Lorentz transformations. 
Moreover, by choosing $\Phi^i$ and the tensor $ \Lambda_{i\bar{r}j\bar{n}} $ to have vanishing Weyl weights,  i.e. 
\bea
\label{weyl-superfields}
\delta_{\Sigma}{\Lambda}_{i\bar{r}j\bar{n}} = \delta_{\Sigma} {\Phi}^i = \delta_{\Sigma} \bar{\Phi}^{\bar{r}} =0,
\eea
and by using (\ref{weyl-superderiv}), one may straightforwardly check that under a super-Weyl transformation,
 the vector superfield (\ref{thevector}), 
 scales as
\bea
\label{weyl-thevector}
\delta_{\Sigma} (\Lambda_{i\bar{r}j\bar{n}} \bar{{\cal{D}}}_{\dot{\alpha}} {\bar{\Phi}}^{\bar{r}} {\cal{D}}_{\alpha} \Phi^i  \bar{{\cal{D}}}^{\dot{\alpha}}{\bar{\Phi}}^{\bar{n}} {\cal{D}}^{\alpha} \Phi^j) = 
-2 ( \Sigma + \bar{\Sigma} ) (\Lambda_{i\bar{r}j\bar{n}} \bar{{\cal{D}}}_{\dot{\alpha}} {\bar{\Phi}}^{\bar{r}} {\cal{D}}_{\alpha} \Phi^i  \bar{{\cal{D}}}^{\dot{\alpha}}{\bar{\Phi}}^{\bar{n}} {\cal{D}}^{\alpha} \Phi^j).
\eea
Of course, when we perform the super-Weyl rescaling to our Lagrangian (\ref{KSB1}), we have to consider 
the variation of the involved superfields in the new $\Theta$ variables \cite{Wess:1992cp}.

\section{The Emergent Potential}

The main result in this section is to show  that, even in those theories where there is no superpotential,
a scalar potential can still be introduced through higher derivatives. 
A hint that this is indeed the case, emerges from the  fact 
that higher derivative matter couplings are likely to change the scalar potential as soon as the equations of motion for 
the auxiliary fields are solved and plugged-back into the action 
\cite{Cecotti:1985mf,Cecotti:1987sa,fer2,Cecotti:1986jy,Cecotti:1986pk,Krasnikov:1989ry,Sasaki:2012ka,Khoury:2010gb} .

\subsection{A Single Chiral Multiplet with no Superpotential}

In order to illustrate the appearance of a scalar potential when the superpotential is vanishing, 
we need to make the effect of the new  coupling (\ref{KSB-Kahler-inv})  more 
transparent. Towards this target,  we will consider first a theory with only \emph{ one chiral multiplet and no superpotential}. 
We will discuss this case first as it will better illustrate our results. In addition,
absence of a superpotential  may be forced by global symmetries (R-symmetry for example) 
which might forbid its appearance.
In this case, the  Lagrangian (\ref{lang}) 
is explicitly written as
\bea
\label{FK1} 
{\cal{L}}=\int d^2 \Theta\,\, 2 {\cal{E}} \left\{ \Big{(}\bar{{\cal{D}}}\bar{{\cal{D}}} - 8 {\cal{R}}\Big{)} 
\left[\frac{3}{8} e^{-\frac{1}{3}K} +\frac{1}{8} \Lambda\ 
\bar{{\cal{D}}}_{\dot{\alpha}} {\bar{\Phi}} {\cal{D}}_{\alpha} \Phi \bar{{\cal{D}}}^{\dot{\alpha}}{\bar{\Phi}} {\cal{D}}^{\alpha} \Phi \right] \right\} + h.c.
\eea
with $\Lambda$ being an abbreviation for ${\Lambda}_{{\Phi}\bar{{\Phi}}{\Phi}\bar{{\Phi}}}$, a hermitian and 
K\"ahler invariant function of $\Phi$ and $\bar{\Phi}$. 
In component form, the bosonic sector of the Lagrangian (\ref{FK1}) turns out to be 
(after integrating out the auxiliary fields $M$ and $b_{a}$, and subsequent appropriately rescalings)
\bea
\label{FK2} 
e^{-1}{\cal{L}}_{\rm{bos}}&=&
\nonumber -\frac{1}{2} R - g_{A {\bar{A}}} \partial_a A \partial^a \bar{A} + g_{A {\bar{A}}}\ e^{\frac{K}{3}} F\bar{F} \\
&& - 16\ {\cal{U}}\ \left\{
\phantom{\frac{X^X}{X^X}}\!\!\!\!\!\!\!\!\!\!\!
e^{\frac{2K}{3}} (F\bar{F})^2
+\partial_a A \partial^a A \partial_b \bar{A}\partial^b \bar{A}
- 2 e^{\frac{K}{3}} F\bar{F}\partial_a A\partial^a \bar{A} \,\right\}
\eea
where ${\cal{U}}$ is a hermitian K\"ahler invariant function of the scalar field (it is the lowest component of $\Lambda$, eq.(\ref{LL})).
 The equation of motion for $F$ is
\bea
\label{Eq-F} 
\bar{F} \Big{(} g_{A {\bar{A}}}- 32\  {\cal{U}} e^{\frac{K}{3}} F \bar{F}
+ 32\  {\cal{U}} \partial_a A\partial^a \bar{A} \Big{)} =0
\eea
which can be easily solved for
\bea
\label{Sol-F} 
F \bar{F} = e^{\frac{-K}{3}} \left( \frac{g_{A\bar{A}}}{32\ {\cal{U}}} +\partial_a A\partial^a \bar{A} \right).
\eea
The other solution $F=0$ brings us back to the standard supergravity case. 
Two important comments are in order here.  
First, one can see that the condition of supersymmetry breaking is changed. 
The {\it{vev}} of $F$, which is related to $\delta_{susy} \chi$, is no longer connected to the 
derivative of the superpotential. It is rather proportional to the potential itself, 
a fact that is reminiscent of the D-term supersymmetry breaking. This  is not that surprising 
since (\ref{thevector}) is a vector superfield in any case. Moreover, in the vacuum, since $F \bar{F}$ 
is positive, ${\cal{U}}$ has to be \emph{positive} defined. The on-shell Lagrangian is then
\bea
\label{FK3} 
e^{-1}{\cal{L}}_{\rm{bos}}= 
-\frac{1}{2} R 
+ \frac{ (g_{A {\bar{A}}})^2}{64\  {\cal{U}}} 
- 16\  {\cal{U}} \partial_a A \partial^a A \partial_b \bar{A}\partial^b \bar{A}
+ 16\  {\cal{U}} \partial_a A\partial^a \bar{A} \partial_b A\partial^b \bar{A}.
\eea
What has happened here has completely changed the dynamics of the theory. 
The minimal kinematic term for the scalar is lost\footnote{This peculiarity is waived as soon as a second chiral multiplet 
is allowed to interact as we will see later.}, and we are only left with terms strongly resembling  
the k-essence \cite{m1,m2}. Much more interesting is the appearance of an \emph{emerging} scalar potential 
\bea
\label{EP} 
{\cal{V}} = - \frac{1}{64} \frac{ (g_{A {\bar{A}}})^2}{  {\cal{U}}}.
\eea
From the positivity of \ ${\cal{U}}$ we see that the potential (\ref{EP}) is negative defined 
\be
{\cal{V}}<0
\ee
and therefore the theory may only have anti-de Sitter vacua. Another important property of the 
emerging potential is that it is not built from a holomorphic function. 

In order an emerging potential to appear in higher derivative theories with chiral multiplets, 
 two fundamental issues should be satisfied
\begin{enumerate}
 \item Instabilities and ghosts states due to higher derivatives should not appear, and
\item  the auxiliary $F$ should not be propagating in which case it can not be integrated out algebraically
\end{enumerate}
The above issues probably make the interaction (\ref{FK1}) unique  and
to our knowledge there is no other higher derivative coupling that can successfully satisfy the above criteria 
and  give rise to an emerging potential. 
In the framework of new-minimal supergravity, consistent higher derivative terms which  
satsify the above restrictions have been considered 
\cite{ger-far,Germani:2011cv}, 
but no scalar potential emerged in that case. 
Higher derivative interactions have been also studied in \cite{bauman}, 
but in that case instabilities appear (as it leads to higher order equations of motions and thus to Ostrogradski instabilities).
Moreover the auxiliary sector can not be integrated out. 
Finally, it should be noted that there is work done in higher derivative matter couplings in the context of 
conformal supergravity as well, see for example \cite{deWit:2010za}. 

\subsection{A Single Chiral Multiplet with Superpotential }

Let us now consider the same single chiral theory but now with a non-trivial superpotential. In this case, the superspace
Lagrangian will explicitly be written as
\bea
\label{FK1-app} 
{\cal{L}}_P=\int d^2 \Theta\,\, 2 {\cal{E}} \left\{ \Big{(}\bar{{\cal{D}}}\bar{{\cal{D}}} - 8 {\cal{R}}\Big{)}
 \left[\frac{3}{8} e^{-\frac{1}{3}K} +\frac{1}{8} \Lambda\ 
\bar{{\cal{D}}}_{\dot{\alpha}} {\bar{\Phi}} {\cal{D}}_{\alpha} \Phi \bar{{\cal{D}}}^{\dot{\alpha}}{\bar{\Phi}} 
{\cal{D}}^{\alpha} \Phi \right] + P(\Phi) \right\} + h.c.
\eea
In component form,  after integrating out 
the auxiliary fields $M$ and $b_{a}$ and performing appropriate rescalings, the bosonic sector of the Lagrangian (\ref{FK1-app}) 
turns out to be
\bea
\label{FK2-app} 
e^{-1}{\cal{L}}_{P}^{\rm{bos}}&=&
\nonumber -\frac{1}{2} R - g_{A {\bar{A}}} \partial_a A \partial^a \bar{A} + g_{A {\bar{A}}}\ e^{\frac{K}{3}} F\bar{F} \\
&&+e^{\frac{2K}{3}} \Big{(} PFK_{A} + \bar{P}\bar{F}K_{\bar{A}} +P_{A}F + \bar{P}_{\bar{A}}\bar{F} \Big{)} +
3e^{K}P\bar{P}\\
&&\nonumber - 16\  {\cal{U}}\ \Big{\{}e^{\frac{2K}{3}} \big{(}F\bar{F}\big{)}^2
+\partial_a A \partial^a A \partial_b \bar{A}\partial^b \bar{A}
- 2 e^{\frac{K}{3}} F\bar{F}\partial_a A\partial^a \bar{A} \Big{\}}.
\eea
In order to extract the on-shell theory we should eliminate the rest of the non-dynamical degrees of freedom, 
a non-trivial procedure as we shall see. The equation of motion for $F$ 
reads
\bea
\label{Eq1-F-app} 
\bar{F} \Big{\{} g_{A {\bar{A}}}- 32 \ {\cal{U}} e^{\frac{K}{3}} \big{(}F \bar{F}\big{)}
+ 32 \ {\cal{U}} \partial^a A\partial^a \bar{A} \Big{\}} =-e^{\frac{K}{3}} \Big{(} PK_{A}  +P_{A} \Big{)}.
\eea
Let us now  define 
\bea
\label{abcx-app}
\nonumber
&a&=-32 \ {\cal{U}} e^{\frac{K}{3}},\\ 
\nonumber
&b&=g_{A\bar{A}}+32\ {\cal{U}}\partial^a A\partial^a \bar{A},\\
&c&=e^{\frac{K}{3}}\Big{|}PK_{A}  +P_{A}\Big{|},\\
\nonumber
&x&=F\bar{F}
\eea
in terms of which equation (\ref{Eq1-F-app}), after being multiplied by its hermitian conjugate, turns out to be
\beq
\label{eq3-F-app}
x(ax+b)^2=c^2.
\eeq
Using Cardano's method, the solutions for $x_{(i)}$ ($i=0, 1 ,2$), are
\bea
\label{cardano-app}
\nonumber
&x_{(i)}&= -\frac{2b}{3a} + t_{(i)},\\ 
\nonumber
&t_{(i)}&=\omega^i \left( -\frac{q}{2} - \sqrt{\frac{q^2}{4}+\frac{p^3}{27}} \right)^{1/3}
-\omega^{3-i} \  \frac{p}{3}  \left( -\frac{q}{2} - \sqrt{\frac{q^2}{4}+\frac{p^3}{27}} \right)^{-1/3}, \\
&\omega&=-\frac{1}{2} +i \frac{\sqrt{3}}{2}
\eea
with
\bea
\label{pq-app}
\nonumber
&p&=-\frac{b^2}{3a^2},\\
&q&=-\frac{2b^3}{27a^3}-\frac{c^2}{a^2}.
\eea
Therefore, the on-shell Lagrangian in terms of the $t_i$,  will have the form
\bea
\label{FK3-app} 
\nonumber e^{-1}{\cal{L}}_{P}^{\rm{bos}}|_{t_i}=&&-\frac{1}{2} R - g_{A {\bar{A}}} \partial_a A \partial^a \bar{A} - 16\  {\cal{U}} \partial_a A \partial^a A \partial_b \bar{A}\partial^b \bar{A}\\ && +3e^{K}P\bar{P} -e^{\frac{K}{3}} \left[ \frac{3}{2} a (t_i)^2 -b t_i\right]
\eea
and supersymmetry is broken for 
\be
\langle x_{(i)}\rangle =\langle-\frac{2b}{3a} +t_{(i)} \rangle \ne 0.
\ee
 These solutions correspond to three different perturbative vacua of (\ref{FK2-app}).
The standard road is to plug $x_0$ into the theory. This corresponds to 
the ordinary vacuum of minimal supergravity, as has been argued by recent work in 
global supersymmetry \cite{Khoury:2010gb,Sasaki:2012ka}. This can be seen first of all by the supersymmetry breaking signal, which remains $P K_A +P_A \ne 0$. A more extensive treatment on this can be found in a resent work \cite{Sasaki:2012ka} in the context of supersymmetry and in \cite{LO} in the context of supergravity.

\section{Gauge Invariant Models}

The Lagrangian (\ref{FK1}) can be straightforwardly be generalized to include gauge invariant 
interactions \cite{Ferrara:1974pu,Cremmer:1982en,Cremmer:1982wb,Bagger:1982ab}. In this case,  
the gauge invariant superspace  Lagrangian is
\bea
\label{FKt-sups}
\nonumber
&&{\cal{L}}_{tot}=\int d^2 \Theta \,\,2 {\cal{E}} \left\{
\frac{3}{8} \Big{(} \bar{{\cal{D}}}\bar{{\cal{D}}} - 8 {\cal{R}} \Big{)} e^{ - {\tilde{K}}/3 } +\frac{1}{16g^2} H_{(ab)}(\Phi) W^{(a)} W^{(b)} 
+ P(\Phi)\right. \nonumber  \\
&&\hspace{3cm}\left.+ \frac{1}{8} 
\Big{(} \bar{{\cal{D}}}\bar{{\cal{D}}} - 8 {\cal{R}} \Big{)} \left[ 
{\tilde{\Lambda}}^{\bar{r}i\bar{n}j}
\ 
\bar{{\cal{D}}}_{\dot{\alpha}} {\tilde{K}}_{i} 
{\cal{D}}_{\alpha}  {\tilde{K}}_{\bar{r}} 
\bar{{\cal{D}}}^{\dot{\alpha}} {\tilde{K}}_{j} 
{\cal{D}}^{\alpha}  {\tilde{K}}_{\bar{n}}  
\right] \right\}+ h.c.
\eea
where
\beq
\label{Kahler-tilde}
{\tilde{K}}=K(\Phi,\bar{\Phi}) + \Gamma(\Phi,\bar{\Phi},V),
\eeq
and
\beq
\label{Gamma-WZgauge}
\Gamma(\Phi,\bar{\Phi},V)=V^{(a)}{\cal{D}}^{(a)} + \frac{1}{2} g_{i {\bar{r}}}  X^{i(a)} {\bar{X}}^{\bar{r}(b)} V^{(a)} V^{(b)}.
\eeq
In addition, as usual, $V^{(a)}$ is the supersymmetric Yang-Mills vector multiplet and 
\be
W_{\alpha}=W_{\alpha}^{(a)}T^{(a)}=-\frac{1}{4}\Big{(} \bar{{\cal{D}}}\bar{{\cal{D}}} - 8 {\cal{R}} \Big{)} e^{-V}{\cal{D}}_{\alpha}e^{V}
\ee
is the gauge invariant chiral superfield containing the gauge field strength. The holomorphic function $ H_{(ab)} $
 is included for generality, but in what follows we will consider $ H_{(ab)} = \delta_{(ab)} $. Expression (\ref{Gamma-WZgauge}) is calculated in the Wess-Zumino gauge, ${\cal{D}}^{(a)}$ 
are the so-called Killing potentials whereas  $X^{i(a)}$ and ${\bar{X}}^{\bar{r}(b)}$ are the components of the 
holomorphic Killing vectors that generate the isometries of the K\"ahler manifold. 
The Killing vectors and the Killing potential are connected via
\bea
\label{X-D}
g_{i {\bar{r}}}{\bar{X}}^{\bar{r}(a)}=i\frac{\partial}{\partial a^i}{\cal{D}}^{(a)},\\
g_{i {\bar{r}}}X^{i(a)}=-i\frac{\partial}{\partial {\bar{a}}^{\bar{r}}}{\cal{D}}^{(a)}
\eea
where $a^i$ and ${\bar{a}}^{\bar{r}}$ are the K\"ahler space complex co-ordinates. We note that the ${\cal{D}}^{(a)}$ that correspond to some $U(1)$ gauged symmetry are only determined up to a constant 
$\xi$, which is the analog for the Fayet-Iliopoulos D-term in supergravity. Now ${\tilde{\Lambda}}^{\bar{r}i\bar{n}j} $ 
has to respect all the isometries of the K\"ahler manifold. 
Again, following the standard procedure,  the bosonic part of the Lagrangian (\ref{FKt-sups}) 
turns out to be
\bea
\label{FKt-comp}
\nonumber
e^{-1}{\cal{L}}_{tot}= && -\frac{1}{2} R - g_{i {\bar{r}}} {\tilde{D}}_m A^i {\tilde{D}}^{m} \bar{A}^{\bar{r}}
 +e^{\frac{K}{3}} g_{i {\bar{r}}} F^i  \bar{F}^{\bar{r}} \\ 
\nonumber
&&  -\frac{1}{16g^2} F_{mn}^{(a)} F^{mn(a)}    -\frac{1}{2} g^2 \big{(} {\cal{D}}^{(a)} \big{)}^2        \\
&& - \ e^{\frac{2K}{3}} \Big{(} F^i D_i P + \bar{F}^{\bar{r}} D_{\bar{r}} \bar{P}\Big{)}   + 3 e^K P \bar{P} \\ 
\nonumber
&&-16\ \tilde{{\cal{U}}}_{i\bar{r}j\bar{n}} \Big{(}e^{\frac{2K}{3}}F^{i} F^{j}\bar{F}^{\bar{r}}\bar{F}^{\bar{n}}
+{\tilde{D}}_a A^{i} {\tilde{D}}^{a} A^{j} {\tilde{D}}_{b} \bar{A}^{\bar{r}} {\tilde{D}}^{b} \bar{A}^{\bar{n}} \\
\nonumber
&&\ \ \ \ \ \ \ \ \ \ -e^{\frac{K}{3}}F^{i} \bar{F}^{\bar{r}} {\tilde{D}}_a A^{j}{\tilde{D}}^{a} \bar{A}^{\bar{n}}
-e^{\frac{K}{3}}F^{i} \bar{F}^{\bar{n}} {\tilde{D}}_a A^{j}{\tilde{D}}^{a} \bar{A}^{\bar{r}}\Big{)}.
\eea
We note that  
\be
{\tilde{D}}_{c} A^{j} =\partial_c A^{j} - \frac{1}{2} B_{c}^{(a)} X^{j}_{(a)} 
\ee
is the covariant derivative and $B_{c}^{(a)}$ is a vector field (belonging to the $V^{(a)}$ 
vector multiplet) that corresponds to the gauged isometries, with 
 field strength $F_{mn}^{(a)}$. 

\subsection{Emergent Potential for a Single Chiral Multiplet with no Superpotential }

In order to illustrate the properties of the {\it{emergent potential}} in the case of gauged models, 
our first example will be a single chiral multiplet with \emph{no} superpotential. In this case the Lagrangian (\ref{FKt-comp}) is
\bea
\label{FKt-comp-single}
\nonumber
e^{-1}{\cal{L}}_{tot}= && -\frac{1}{2} R - g_{A {\bar{A}}} {\tilde{D}}_m A {\tilde{D}}^{m} \bar{A}
 +e^{\frac{K}{3}} g_{A {\bar{A}}} F  \bar{F} \\ 
&&  -\frac{1}{16g^2} F_{mn}^{(a)} F^{mn(a)}    -\frac{1}{2} g^2 \big{(} {\cal{D}}^{(a)} \big{)}^2        \\
\nonumber
&&-16\ \tilde{{\cal{U}}} \Big{(}e^{\frac{2K}{3}} (F\bar{F})^2
+{\tilde{D}}_a A {\tilde{D}}^{a} A {\tilde{D}}_{b} \bar{A} {\tilde{D}}^{b} \bar{A} - 2\ e^{\frac{K}{3}}F 
\bar{F} {\tilde{D}}_a A{\tilde{D}}^{a} \bar{A} \Big{)}.
\eea
The single auxiliary field $F$ can now be eliminated from (\ref{FKt-comp-single}) by its equations of motion, leading to 
\bea
\label{Sol-FG} 
F \bar{F} = e^{\frac{-K}{3}} \left( \frac{g_{A\bar{A}}}{32\  \tilde{{\cal{U}}}} +{\tilde{D}}_a A {\tilde{D}}^a \bar{A} \right).
\eea
Plugging (\ref{Sol-FG}) back in (\ref{FKt-comp-single}), we can easily read-off the potential for the gauged model
which turns out to be
\bea
\label{UpliftedEP} 
{\cal{V}} = \frac{1}{2} g^2 \left({\cal{D}}^{(a)} \right)^2- \frac{(g_{A\bar{A}})^2}{64\  \tilde{{\cal{U}}}}
\eea
with $\tilde{{\cal{U}}}=\tilde{{\cal{U}}}_{A\bar{A}A\bar{A}}$, a K\"ahler-space 
tensor that respects all the isometries of the gauged group. 
For a first example we will take a flat model with K\"ahler potential 
\be
K=a \bar{a} +d\, 
\ee
which leads to
 \be
g_{a\bar{a}}=1\, , ~~~~{\cal{R}}_{a\bar{a}a\bar{a}}=0
\ee
The   $U(1)$ Killing potential is
\be
 D^{(1)}=a \bar{a} + \xi
\ee
where the parameter $\xi$ corresponds to the aforementioned freedom to 
shift the $U(1)$ Killing potential. When we promote $a$ and $\bar{a}$ to the superfields $\Phi$ and $\bar{\Phi}$,
 our K\"ahler potential $K$ together with the counter term $\Gamma$ become
\beq
\label{Kahler-tilde-U(1)}
{\tilde{K}}_{U(1)}=\Phi \bar{\Phi} + V \Phi \bar{\Phi} +\frac{1}{2} V^2 \Phi \bar{\Phi} +d +V \xi.
\eeq
The bosonic part of our Lagrangian in component form then turns out to be
\bea
\label{FKt-comp-U(1)}
\nonumber
e^{-1}{\cal{L}}_{U(1)}= && -\frac{1}{2} R -\frac{1}{16g^2} F_{cd} F^{cd} \\
&&-16\ \tilde{{\cal{U}}} {\tilde{D}}_a A {\tilde{D}}^{a} A {\tilde{D}}_{b} \bar{A} {\tilde{D}}^{b} \bar{A} + 16\  \tilde{{\cal{U}}} {\tilde{D}}_a A {\tilde{D}}^{a} \bar{A} {\tilde{D}}_{b} A {\tilde{D}}^{b} \bar{A}  \\ \nonumber
&&-\frac{1}{2} g^2 \left( A \bar{A} + \xi  \right)^2 + \frac{1}{64\  \tilde{{\cal{U}}}},
\eea
with ${\tilde{D}}_{m} A =\partial_m A + \frac{i}{2} B_{m} A $. Then the  scalar potential is
\bea
\label{EP-U(1)} 
{\cal{V}} = \frac{1}{2} g^2 \big{(} D^{(a)} \big{)}^2- \frac{1}{64 \, \tilde{{\cal{U}}}}.
\eea
A simple choice for $\tilde{{\cal{U}}}$ could  be 
\be
\tilde{{\cal{U}}}=mg_{A\bar{A}}g_{A\bar{A}}=m\, ,
\ee
 where $m$ is a positive constant. 
It is also possible to allow $m$ to be some function of the Killing potential 
\be
D^{(1)}=a \bar{a} + \xi
\ee
as we will see  in a moment. 
From (\ref{EP-U(1)}) one can see that the interplay between $\xi$, $g$ and $m$ provides de Sitter, Minkowski or anti-de Sitter 
vacua, all with broken supersymmetry
\bea 
\nonumber
&1)&\text{anti-de Sitter}:\ \ \frac{1}{2}g^2\xi^2<\frac{1}{64m}\\
\nonumber
&2)&\text{Minkowski}:\ \ \ \ \frac{1}{2}g^2\xi^2=\frac{1}{64m}\\
\nonumber
&3)&\text{de Sitter}:\ \ \ \ \ \ \ \ \frac{1}{2}g^2\xi^2>\frac{1}{64m}.
\eea
Other choices of $L$ are also possible. For example by choosing 
\be
\tilde{{\cal{U}}}=\frac{k}{64}\frac{(A\bar{A})^2}{A\bar{A}+\xi} \label{ll1}
\ee 
a richer structure for the potential emerges depending on the values of the parameter $k$. The 
shape of the potential  (\ref{EP-U(1)}) for various values of the parameter $k$, in units of $g^2\xi$ is plotted in Fig.1. 
We see that depending on the value of $k$, we may have stable de Sitter  (branches III,IV), Minkowski (branch II) or
anti-de Sitter backgrounds (branch I). A general property is nevertheless the appearance of metastable de Sitter backgrounds 
for a large range of $k$. 

\begin{figure}[htp] \centering{
\includegraphics[scale=0.82]{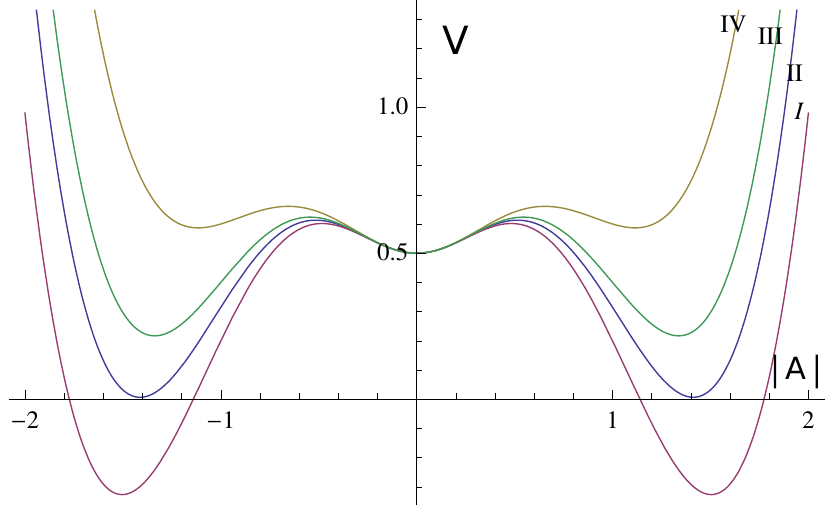}}
\caption{The shape of the scalar potential (\ref{EP-U(1)}) in units of $g^2 \xi$
 for $\tilde{L}$ given in (\ref{ll1})  for various values of the parameter $k$. I. $k=3.6 \, \xi$,~ II. $k=3.37\, \xi$,~ III. $k=3.2\,
\xi$,~ IV. $k=2.8\,\xi$  }
\end{figure}
 
As a second example, we can take 
\be
K=\ln(1+a \bar{a})
\ee
 for the K\"ahler potential, and 
\bea
D^{(1)}= \frac{1}{2} \frac{a+\bar{a}}{(1+ a \bar{a})},\ \ 
D^{(2)}= -\frac{i}{2} \frac{a-\bar{a}}{(1+ a \bar{a})},\ \ 
D^{(3)}= -\frac{1}{2} \left( \frac{1-a\bar{a}}{1+ a \bar{a}} \right)
\eea
for the $S^2=SU(2)/U(1)$ Killing potentials. In this case we get that
\be
g_{a\bar{a}}=\frac{1}{(1+ a \bar{a})^2} \, , ~~~~
{\cal{R}}_{a\bar{a}a\bar{a}}=-\frac{2}{(1+ a \bar{a})^4}
\ee. 
Then, for a constant positive parameter $m$ and  
\be
\tilde{{\cal{U}}}=mg_{A\bar{A}}g_{A\bar{A}},
\ee 
by using that 
\be
\left( {\cal{D}}^{(1)} \right)^2+ 
\left( {\cal{D}}^{(2)} \right)^2+\left( {\cal{D}}^{(3)} \right)^2=\frac{1}{2}
\ee
we find that the scalar potential turns out to be
\bea
\label{EP-SU(2)/U(1)=S^2} 
{\cal{V}}_{S^2,EP} = \frac{1}{4} g^2 - \frac{1}{64 m}\, .
\eea
From (\ref{EP-SU(2)/U(1)=S^2}) we see that the only effect of the uplifted emerging 
potential is to inherit the theory with some cosmological constant, depending on the parameters $\xi$, $g$ and $m$.

\subsection{Emergent Potential for Two Chiral Multiplets with no Superpotential }

We now introduce a second chiral multiplet, but still {\it{no}} superpotential.
The only restriction we place on the K\"ahler potential is that it has the form
\beq
\label{2ChiralK}
K=K_1(\Phi_1,\bar{\Phi}_1)+K_2(\Phi_2,\bar{\Phi}_2)+d
\eeq
with $K_1$ and $K_2$ hermitian functions of $\Phi_1,\bar{\Phi}_1$ and $\Phi_2,\bar{\Phi}_2$, respectively. 
Moreover we shall restrict the higher derivative scalar couplings to the simple form 
\be
{\tilde{\Lambda}}_{i\bar{r}j\bar{n}}= m \tilde{K}_{i\bar{r}}\tilde{K}_{j\bar{n}}\, , 
\ee
 with $m$ some positive definite function of the superfields that respects the gauged isometries. 
The equations for the two auxiliaries $F^1$ and $F^2$ are
\bea
\label{2ChiralF}
\nonumber
16g_{2\bar{2}}\ y\ {\tilde{D}}_a A^2 {\tilde{D}}^{a} \bar{A}^{\bar{1}}\!-\!32e^{\frac{K}{3}}x
\!+\!16g_{j\bar{n}}{\tilde{D}}_a A^j {\tilde{D}}^{a} \bar{A}^{\bar{n}} 
\!+\!16g_{1\bar{1}}{\tilde{D}}_a A^1 {\tilde{D}}^{a} \bar{A}^{\bar{1}}
=-m^{-1},
\\
16g_{1\bar{1}}\ \frac{1}{y}\ {\tilde{D}}_a A^1 {\tilde{D}}^{a} \bar{A}^{\bar{2}}\!-\!32e^{\frac{K}{3}}x
\!+\!16g_{j\bar{n}}{\tilde{D}}_a A^j {\tilde{D}}^{a} \bar{A}^{\bar{n}} 
\!+\!16g_{2\bar{2}}{\tilde{D}}_a A^2 {\tilde{D}}^{a} \bar{A}^{\bar{2}}
=-m^{-1}
\eea
where we have defined
\be
x=g_{j\bar{n}}F^j\bar{F}^{\bar{n}}\, , ~~~ y=\frac{\bar{F}^{\bar{2}}}{\bar{F}^{\bar{1}}}.
\ee
 Equations (\ref{2ChiralF}) can be combined in order to provide a solution for $x$ which is found to be
\bea
\label{x}
\nonumber
x=&&\frac{1}{4} e^{\frac{-K}{3}} \left( \frac{1}{8m} +3 g_{j\bar{n}}{\tilde{D}}_a A^j {\tilde{D}}^{a} \bar{A}^{\bar{n}}\right) \\ 
&&\pm \frac{1}{4} e^{\frac{-K}{3}} \left\{ \Big{(}g_{1\bar{1}}{\tilde{D}}_a A^1 
{\tilde{D}}^{a} \bar{A}^{\bar{1}}-g_{2\bar{2}}{\tilde{D}}_a A^2 {\tilde{D}}^{a} \bar{A}^{\bar{2}}\Big{)}^2\right. \\ 
\nonumber
&&\left.\ \ \ \ \ \ \ \ \ \ \ \ + 4g_{1\bar{1}}g_{2\bar{2}}{\tilde{D}}_a A^2 {\tilde{D}}^{a} 
\bar{A}^{\bar{1}}{\tilde{D}}_b A^1 {\tilde{D}}^{b} \bar{A}^{\bar{2}} \right\}^{\frac{1}{2}}.
\eea
Then, plugging back into the action (\ref{FKt-comp}), the on-shell theory is
\bea
\label{FK4}
\nonumber
e^{-1}{\cal{L}}_{tot}= && -\frac{1}{2} R - \frac{1}{4} g_{i {\bar{r}}} {\tilde{D}}_a A^i {\tilde{D}}^{a} \bar{A}^{\bar{r}} -\frac{1}{16g^2} F_{cd}^{(a)} F^{cd(a)} +\frac{1}{64m}  -\frac{1}{2} g^2 \left( {\cal{D}}^{(a)} \right)^2   \\ 
\nonumber
&&+ 9m \Big{(} g_{i {\bar{r}}} {\tilde{D}}_a A^i {\tilde{D}}^{a} \bar{A}^{\bar{r}} \Big{)}^2 -16mg_{i {\bar{r}}} g_{j {\bar{n}}}  {\tilde{D}}_a A^{i} {\tilde{D}}^{a} A^{j} {\tilde{D}}_{b} \bar{A}^{\bar{r}} {\tilde{D}}^{b} \bar{A}^{\bar{n}} \\
\nonumber
&&+4m g_{1\bar{1}}g_{2\bar{2}}{\tilde{D}}_a A^2 {\tilde{D}}^{a} \bar{A}^{\bar{1}}{\tilde{D}}_b A^1 {\tilde{D}}^{b} 
\bar{A}^{\bar{2}} +m \Big{(}g_{1\bar{1}}{\tilde{D}}_a A^1 {\tilde{D}}^{a} 
\bar{A}^{\bar{1}}-g_{2\bar{2}}{\tilde{D}}_a A^2 {\tilde{D}}^{a} \bar{A}^{\bar{2}}\Big{)}^2 \\ 
\nonumber
&& \pm \Big{(} \frac{1}{4} + 6 m g_{i {\bar{r}}} {\tilde{D}}_a A^i {\tilde{D}}^{a} \bar{A}^{\bar{r}}\Big{)}
 \left\{ \Big{(}g_{1\bar{1}}{\tilde{D}}_a A^1 {\tilde{D}}^{a} \bar{A}^{\bar{1}}-g_{2\bar{2}}{\tilde{D}}_a A^2 
{\tilde{D}}^{a} \bar{A}^{\bar{2}}\Big{)}^2\right. \\ 
&&\left.
\!\!\!
\phantom{\Big{)}^2} 
 + 4g_{1\bar{1}}g_{2\bar{2}}{\tilde{D}}_a A^2 {\tilde{D}}^{a} \bar{A}^{\bar{1}}{\tilde{D}}_b A^1 
{\tilde{D}}^{b} \bar{A}^{\bar{2}} \right\}^{\frac{1}{2}}
\eea
where we notice that the minimal kinetic terms have not dissapeared and a number of DBI-like 
kinetic terms have appeared in the action along with the higher derivatives. 
The interchanging $\pm$ signs inside the Lagrangian are a manifestation of the higher 
derivative (safe nevertheless) nature of this supersymmetric theory, due to which, 
there exists the possibility to have more than one solutions for the auxiliary fields. 
The interesting part is that the potential of this theory is still the uplifted emergent potential 
(\ref{UpliftedEP}), thus, with a suitable choice of the K\"ahler and Killing potentials one can achieve 
the various properties discussed earlier. For example, for a $U(1)$ gauged isometry we have 
\be
K=a_1 \bar{a}_{\bar{1}}+a_2 \bar{a}_{\bar{2}} +d\, , ~~~~D^{(1)}=a_1 \bar{a}_{\bar{1}}+a_2 \bar{a}_{\bar{2}} 
 + \xi\, ,
\ee
 with the scalar potential given by
\bea
\label{2ChiralUpliftedEP} 
{\cal{V}} = \frac{1}{2} g^2 \left(A_1 \bar{A}_{\bar{1}}+A_2 \bar{A}_{\bar{2}}  + \xi \right)^2- \frac{1}{64 m}
\eea
where $m$ can be a positive constant or a gauge invariant positive definite hermitian function of the 
fields $A^1,\bar{A}^{\bar{1}}$ and $A^2,\bar{A}^{\bar{2}}$.

Summarizing,   the well-known standard form of the ${\cal{N}}=1$ potential (\ref{poten}) 
is only valid at the leading two-derivative 
level. Whenever higher-derivatives are introduced, an emerging scalar potential appear even when there is no superpotential.
The emerging potential is negative defined and can be uplifted to positive values in gauge chiral models by  D-term contributions. 
There are many open problems to be discussed in the future among which is the possible applications of our findings 
to High Energy phenomenology.



 \begin{acknowledgments}
The authors thank C. Germani, U. Lindstr\"om and G. Orfanidis for discussion. While this work was in the final stage, another work by M. 
Koehn, J.-L Lehners and B. Ovrut \cite{LO} appeared with partial overlapping material in the non-gauged models.

\end{acknowledgments}


\end{document}